\documentclass [12pt, titlepage, a4paper]{article}
\usepackage[latin1]{inputenc}
\usepackage[german,english]{babel}
\usepackage{amsmath, amssymb, graphicx, latexsym}

\begin{document}

\thispagestyle{empty}
\begin{flushright}
IPPP/08/87 \\
DCPT/08/174
\end{flushright}
\vspace{0.5cm}
\begin{center}
\Large{Numerical evidence for loop convergence in\\
Yang-Mills thermodynamics}

\vspace{2cm}

\large{Dariush Kaviani}

\vspace{1.5cm}

\small{\textit{Institute for Partical Physics Phenomenology\\
Ogden Center for Fundamental Physics\\
Durham University\\
South Road\\
DH1 3LE Durham, UK}}

\vspace{2.5cm}

\end{center}

\thispagestyle{empty}

\begin{center}
\parbox{15cm}{
\begin{abstract}
The numerical results for the computed moduli of the irreducible three-loop contributions to the thermodynamical pressure of an SU(2) Yang-Mills theory in the effective theory for the deconfining phase are explained in detail. Irreducible three-loop integrations are compared with two-loop integrations and the different nature of their integrations is scrutinized and illustrated numerically. The numerical results show a rapid convergence in the loop expansion of Yang-Mills thermodynamics. The statistical method used for irreducible three-loop integrations is explained and checked for two-loop integrations. The statistical results for two-loop integrations are compatible with the former computed analytical results showing the reliability of the statistical method. This is a companion paper to [1].
\end{abstract}
\selectlanguage{english}
}
\end{center}
\thispagestyle{empty}
\cleardoublepage

\section{Introduction}
One of the major problems in constructing Yang-Mills thermodynamics is that a reliable approximation of high-temperature thermodynamics related to 4D Yang-Mills theories in terms of the small-coupling expansion seems to be impossible.

The nonconvergence of the small-coupling expansion is due to the fact that an empty (trivial) ground-state is invoked to construct an approximating series for the full partition function. Meanwhile, fluctuations of nontrivial topology have a profound impact on the ground-state estimate, and are completely ignored in small-coupling expansion since their weight posesses an essential zero at vanishing coupling.

Consequently, the strong correlating effects of these extended field configuratons are completely ignored. This is a fact which is expressed by tree-level masslessness and only week radiative screenings of all gauge bosons leading to the nonconvergence of the expansion. However, by considering an a priori estimate for the ground state of an SU(2) Yang-Mills theory at high temperatures, which is obtained by a self-consistent and sufficiantly local spatial coarse graining over interacting and stable BPS saturated topological field configurations, according to the effective theory reviewed in \cite{1}, leads to a rapidly converging loop expansion \cite{1,2}.

 The argument of \cite{2} is that a self-consistent spatial coarse-graining, which involves interacting (anti)calorons of unite topological charge moulus, implies that real time loop expansions of thermodynamical quantities in the deconfining phase of SU(2) and SU(3) Yang-Mills thermodynamics are, IPI resummations, determined by a finite number of connected bubble diagrams.
 
To see this in more detail, recall the following fundamental aspects of the effective theory. In the effective theory a composite adjoint Higgs field $\phi$ which describes the topologically nontrivial part of the ground state is introduced and implimented. This field is associated with periodic instantons (Calorons of topological charge one and trvial holonomy) of high temperature in the deconfining phase and is used as a background for coarse-grained topologically nontrivial sector of the theory. This field is quantum mechanically and thermodynamically stabilized. 

A macroscopic pure-gauge configuration which is the solution of the equation of motion for the topologically trivial sector in the presence of the background is implimented in order to include the interactions between trivial holonomy Calorons.

As the modulus of the Higgs field decreases with the temperature as \mbox{$|\phi| \sim \sqrt{\frac{\Lambda^3}{T}}$}, where $\Lambda$ is the Yang-Mills scale, the effects of topological defects die off at large temperature in a power-like fashion. Asymtotic freedom and Infrared-ultraviolet decoupling of the fundamental thery, which are the results obtained in perturbation theory at zero temperature, are preserved.

Thermodynamical quantities are in this framework calculated as loop expansions about the nontrivial ground state consisting of the Higgs field and the pure gauge configuration. The tree-level excitations are either massive thermal quasiparticles or massless photons, which interact very weakly. The effective theory is both infrared-and ultaviolet finite. The former property is due to the existence of Caloron-induced gauge boson masses (IR cut-off), and the later to the compositeness constraints on the loop momenta arising from the existense of the composite scale of the Higgs field (UV cut-off).

By taking the topologically nontrivial contributions through the effective theory for deconfining SU(2) Yang-Mills thermodynamics into account, the composite Higgs field induces a dynamical gauge symmetry breaking from SU(2) $\rightarrow$ U(1) implying that two out of the three propagating and coarse-grained gauge modes acquire temperature dependent mass. According to this and the fact that the off-shellness of these modes, along with the momentum transfer in local vertices, are highly constrained by spetial coarse-graining a rapid convergence the loop expansion is expected \cite{1,2}.

The aim of this paper is to provide numrical evidence for the convergence in the loop expansion of Yang-Mills thermodynamics \cite{1}. It is worth emphasising that this paper contains illuminating details and some technical aspects not present in \cite{1}. 

The paper is organized as follows. Section 2 considers the effective gauge coupling, which lies at the heart of the numerical results, and the relating Feynman-rules with the constraints on the momenta for loop integrations are also explained. These were not displayed explicitly in \cite{1}.
Section \ref{chp:constraintsandcompactness} compares irreducible three-loop diagrams with two-loop diagrams, arguing analytically that integrations concerning the former are, in contrast to the later, either compact or empty and therefore give much more suppressed contributions to the thermodynamical pressure. Section 4 gives a mini review of the Monte-Carlo method used for loop integrations and verifies numerically the analytical claim of section 3 by using numerical illustration and Mote-Carlo integration. Numerical illustration shows that the region of radial loop integration for irreducible three-loop contributions is, in contrast with the region of radial loop integration for two-loop, either compact or empty. Monte-Carlo integration shows that irreducible three-loop integrations lead to much more suppressed contributions to the thermodynamical pressure than two-loop integrations. This is precisely in agreement with the result of the analytical argument of section 3. The stability and reliability of the Monte-Carlo method used for doing loop integrations is also tested by comparing the results of the Monte-Carlo integration for two-loop with the former analytical result for two-loop. Section 5 summerizes the objectives achieved and concludes with an open problem.

\section{Effective gauge coupling and Feynman rules } 
In \cite{1} the effective theory was reviewed by taking the effective action as the startig point resulting into an evolution equation $\lambda(a)$ for temperature as a function of tree-level gauge boson mass. This evolution has two fixed points $a = 0$ and $a = \infty$, where the lowest and highest temperatures $\lambda_c = \lambda(a=\infty)$ and $\lambda_p = \lambda(a = 0)$ are attainable, respectively.

The evolution $\lambda(a)$ can be inverted to yield an evolution $e(\lambda)$ for the effective gauge coupling as a function of temperature. In what follows the evolution of $e(\lambda)$ is displayed, since it lies at the heart of the numerical results, and the relating Feynman rules, which depend on the effective gauge coupling, are also written down explicitly. These were used in \cite{1} for doing loop integrations, but were not presented explicitly.  
For SU(2) and SU(3) this is illustrated in figure 2.1. The evolution of e with temperature exhibits a logarithmic pole, $e\propto -\log(\lambda-\lambda_c)$, 
where $\lambda_c = 13.89$ denotes the critical value of the dimensionless temperature $\lambda\equiv\frac{2\pi T}{\Lambda}$ and the 
value of e at the plateau is $e=\sqrt{8}\pi \sim 8.89.$.

\begin{figure}[!ht]
\begin{center}
\includegraphics[scale=0.8]{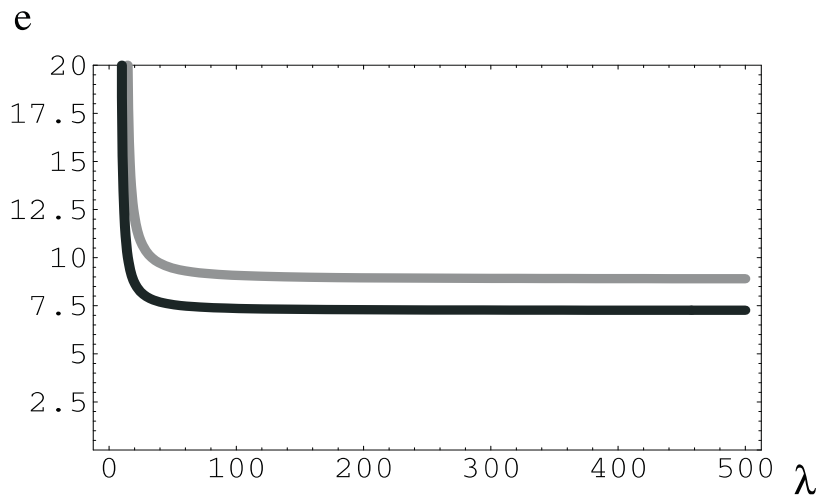}
\caption{Evolution of the effective gauge coupling e in the electric (deconfining) phase for SU(2) (grey line) and SU(3) (black line).}
\end{center}
\end{figure}

In total, the theory seems to have three phases: The electric phase at high temperatures, the magnetic phase for a small range of 
temperatures comparable to the Yang-Mills-scale $\Lambda$ and a center phase for low temperatures. The electric phase is deconfining, 
the magnetic phase is preconfining and the center phase completely confining. Here we are only interested in the deconfining 
(electric) phase.

Now by knowing the effective gauge coupling one can formulate the Feynman rules in the unitary-Coulomb gauge
using the real-time formulation of finite-temperature field theory. The real-time is preferable because the 
implementation of constraints on the momenta ((\ref{2_g10}), (\ref{2_g11}), (\ref{2_g15})) is rather inconvenient in the imaginary-time formalism.
It should be noticed that the unitary-Coulomb gauge is a  completely fixed gauge, therefore no Faddeev-Popov determinants need to be considered and no ghost fields need to be introduced. According to the discussion in the introduction the effective theory has a stabilized, composite and adjoint Higgs field $\phi$ characterizing 
its ground state, where in the unitary gauge $\phi$ is diagonal and the pure-gauge background is zero.

The physical gauge choice for a residual gauge freedom due to the unbroken Abelian subgroup $U(1)$ is the Coulomb gauge.
In the unitary-Coulomb gauge each of the propagators for Tree-Level-Heavy (TLH)/Massless (TLM) modes split into a vacuum and thermal part as follows:


\begin{eqnarray}
\label{2_g8}
D^{TLH}_{\mu\nu,ab}(p) & = & -\delta_{ab}\tilde D_{\mu\nu}\left[\frac{i}{p^2-m^2} + 2\pi\delta(p^2-m^2)n_B\frac{(|p^0|)}{T}\right] \\[1ex]
\tilde D_{\mu\nu} & = & (g_{\mu\nu}-\frac{p_{\mu}p_{\nu}}{m^2}) \notag\\[1ex]
\label{2_g8a}
D^{TLM}_{\mu\nu,ab}(p) &=& -\delta_{ab}\left\{P^{T}_{\mu\nu}\left[\frac{i}{p^2} + 2\pi\delta (p^2)n_B \frac{(|p^0|)}{T}\right] - \frac{iu_{\mu}u_{\nu}}{\vec{p}^2}\right\} \\[1ex]
\label{2_g8b}
P^{00}_T &=& P^{0i}_T = P^{i0}_T = 0\\[1ex]
P^{ij}_T &=& \delta^{ij} - \frac{p^{i}p^{j}}{\vec{p}^2}\notag
\end{eqnarray}

\begin{figure}
\begin{center}
\includegraphics{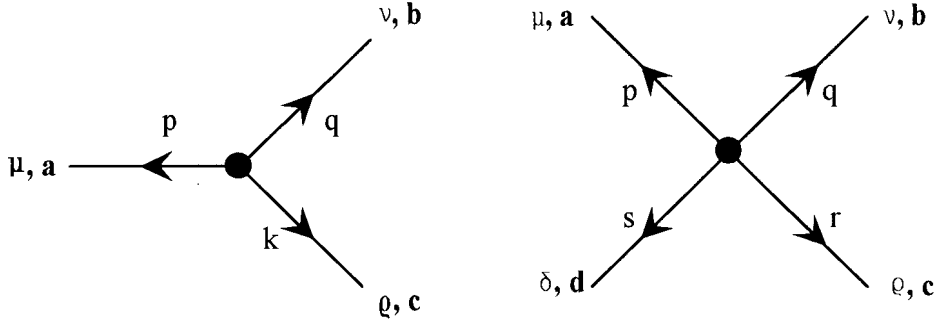}
\caption{Three- and four-vertices with Lorenz and color indices.}
\label{2_f1}
\end{center}
\end{figure}

where $n_B(x)=\frac{1}{e^x-1}$ denotes the Bose-Einstein distribution function. TLM modes carry a color index 3 while TLH 
modes have a color index 1 and 2. It should be noticed that the term $\propto u_{\mu}u_{\nu}$ is due to the "propagation" of 
the $A^3_0$ field, and $u_{\mu} = (1,0,0,0)$ represents the four-velocity of the heat bath \cite{4}.

The three- and  four-vertices for the gauge vector bosons are represented in figure \ref{2_f1} and read:


\begin{eqnarray}
\label{2_g8_9}
\Gamma^{\mu\nu\rho}_{[3]abc} (p,k,q) & = & e(2\pi)^4 \delta (p+q+k) \varepsilon_{abc}[g^{\mu\nu}(q-p)^{\rho}+g^{\nu\rho}(k-q)^{\mu}+ \notag\\[1ex]
&& g^{\rho\mu}(p-k)^{\nu}]   \\[2ex]
\Gamma^{\mu\nu\rho\delta}_{[4]abcd} & = & -ie^2(2\pi)^4\delta(p+q+s+r) \notag \\[1ex]
&&[\varepsilon_{fab}\varepsilon_{fcd}(g^{\mu\rho}g^{\nu\sigma}-g^{\mu\sigma}g^{\nu\rho})+  \\[1ex]
&& \varepsilon_{fac}\varepsilon_{fdb}(g^{\mu\sigma}g^{\rho\nu}-g^{\mu\nu}g^{\rho\sigma})+ \notag \\[1ex]
&& \varepsilon_{fad}\varepsilon_{fbc}(g^{\mu\nu}g^{\sigma\rho}-g^{\mu\rho}g^{\sigma\nu})]. \notag
\end {eqnarray}

The maximal off-shellness of momenta gauge modes due to the nontrivial ground state associated 
with the resolution $|\phi|$ is constrained as:


\begin{equation}
\label{2_g10}
|p^2-m^2|  \leq  |\phi|^2 \quad  \mbox{(TLH modes)}\quad\quad |p^2| \leq  |\phi|^2  \quad \mbox{(TLM modes)}.
\end {equation}

where for TLH modes the mass is given by $m = 2e|\phi| = 2e\sqrt{\frac{\Lambda^3}{2\pi T}}$ and for TLM we have $m=0$.

The other kinematical constraint is on the center-of-mass energy flowing into a four-vertex that should not be greater than the 
compositeness \mbox{scale $|\phi|$} of the effective theory.
For the momenta modes $p$ and $k$ entering the four-vertex one has


\begin{equation}
\label{2_g11}
|(p+k)^2| \leq |\phi|^2.\\[2ex]
\end {equation}

This relation puts a strong restriction on loop integrations. Now consider $(p_1, p_2)$ and $(p_3, p_4)$ 
as the pair of ingoing and outgoing momenta as represented in the diagram below.

\begin{figure}[h]
\begin{center}
\includegraphics{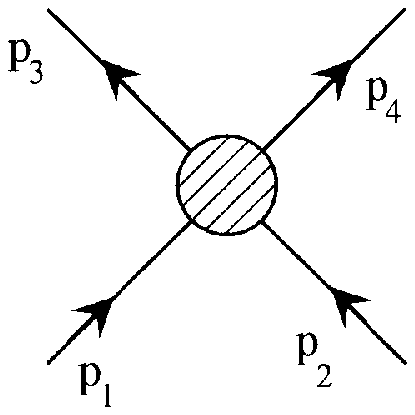}
\end{center}
\label{4_f2}
\end{figure}

The Mandelstam variables are defined as:
\begin {eqnarray}
\label{2_g14}
s &=& (p_1 + p_2)^2 = (p_3 + p_4)^2, \qquad t = (p_3 - p_1)^2 = (p_4 - p_2)^2, \notag \\
u &=& (p_4 - p_1)^2 = (p_3 - p_2)^2. 
\end{eqnarray}

Accordingly, relation (\ref{2_g11}) reads then in terms of s-, t- and u-channels as the following \cite{2,3}:
\begin {eqnarray}
\label{2_g15}
|(p_1 + p_2)^2| &\leq& |\phi|^2 \quad \mbox{(s-channel)}, \quad |(p_3 - p_1)^2| \leq |\phi|^2 \quad \mbox{(t-channel)}, \notag \\
|(p_3 - p_2)^2| &\leq& |\phi|^2 \quad \mbox{(u-channel)}.
\end{eqnarray}

For the three-vertex conditions (\ref{2_g15}) are already contained in (\ref{2_g10}) by momentum 
conservation in the vertex.

From the above conditions one can immediately see the increase in the number of compositeness constraints 
by the s-, t- and u-channels. Conditions (\ref{2_g10}) and (\ref{2_g15}) imply that the higher the loop 
order, the more suppressed their contribution to a thermodynamical quantity. General arguments suggest 
that, apart from diagrams associated with one-particle irreducible resummations of propagators, only 
a finite number of diagrams contribute to the loop expansion \cite{2}. The aim of the present article is 
to demonstrate the numerical evidence of this at three-loop level.

\section{Constraints and Compactness}
\label{chp:constraintsandcompactness}
The following concerns the most important point\footnote{this was first pointed out in \cite{2}} about irreducible three-loop diagrams. In \cite{1} the only irreducible three-loop diagrams were considered,

\begin{figure}[h]
\begin{center}
\includegraphics{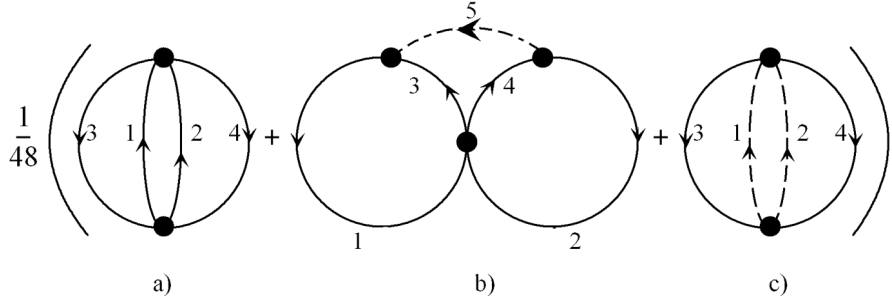}
\caption{Irreducible three-loop contributions to the pressure. Solid (dashed) lines are associated with the propagators of massive (massless) modes.}
\label{4_f1}
\end{center}
\end{figure}

and their moduli contribution to the thermodynamical pressure were computed. For all these irreducible three-loop diagrams we have $\tilde{k} < k$, where \mbox{$\tilde{k} = 6$} is the number of radial independent loop variables $(p^0, |\vec{p}|)_i$ for \mbox{$i \!=\! 1, 2, 3$}, and $k = 7$ is the number of constraints on them, which is counted as follows. For diagram (a) and (b) there are 3 compositeness constraints\footnote{these are used in the next section to verify the supports of radial loop integration - integrations over $(r, \theta, \varphi)$.} emerging from the effective theory over s-, t- and u-channels at the four vertex $|(p_1+p_2)^2| \leq |\phi|^2$, $|(p_3-p_1)^2| \leq |\phi|^2$, $|(p_2-p_3)^2| \leq |\phi|^2$ and 4 on-shellness relations $p_1^2 = m^2$, $p_2^2 = m^2$, $p_3^2 = m^2$ and $p_4^2 =(p_1 + p_2 - p_3)^2 =m^2$; the latter is due to momentum conservation at the four-vertex. Thus, for diagram (a) and (b) we have $k = 3 + 4 = 7$.
Diagram (c) is subject to case differentiation by momentum conservation at the four-vertex\footnote{by using $p_4^2=(p_1+p_2-p_3)^2$, one can find dimensionless equations in terms of radial and angular variables which contradict $p_1^2=p_2^2=0$ and agree with either $p_1^2=0$, $p_2^2\neq0$ or $p_2^2=0$, $p_1^2\neq0$}.
For the case where both of the massless modes propagate off-shell there are 5 compositeness constraints $|(p_1+p_2)^2| \leq |\phi|^2$, $|(p_3-p_1)^2| \leq |\phi|^2$, $|(p_2-p_3)^2| \leq |\phi|^2$, $|p_1^2| \leq |\phi|^2$, $|p_2^2| \leq |\phi|^2$, and 2 on-shellness relations $p_3^2=m^2$, $p_4^2=(p_1+p_2-p_3)^2=m^2$ making $k = 7$.
For the case where one of the massless modes propagate off-shell while the other one is on-shell there are 4 compositeness constraints $|(p_1+p_2)^2| \leq |\phi|^2$, $|(p_3-p_1)^2| \leq |\phi|^2$, $|(p_2-p_3)^2| \leq |\phi|^2$, $|p_{1 or 2}^2| \leq |\phi|^2$, and 3 on-shellness relations $p_3^2=m^2$, $p_4^2 = (p_1+p_2-p_3)^2 = m^2$, $ p_{1 or 2}^2 = 0$ making again $k=7$. Thus, for both cases in diagram (c) we have $k=5+2=4+3=7$, and therefore we conclude that for all the irreducible three-loop diagram $\tilde{k}<k$. The fact $\tilde{k}<k$ shows that irreducible three-loop integrations are either compact or empty \cite{1}, and this is in sharp contrast with the two-loop case. For instance, consider the following nonvanishing two-loop diagrams.

\begin{figure}[!h]
\begin{center}
\includegraphics{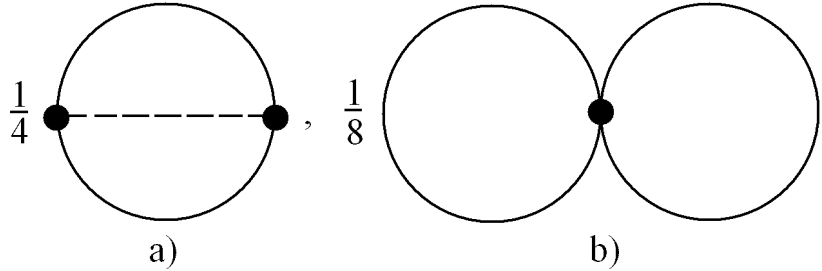}
\caption{Non-local and local two-loop contributions to the pressure.}
\label{3_f2}
\end{center}
\end{figure}

For both of these two-loop diagrams one has $\tilde{k}=4$ since in each case $(p^0, |\vec{p}\;|)_i$ for $i=1,2$.
For both of these diagrams there is one composite constraint $|(p_1+p_2)^2|\leq |\phi|^2$ and two on-shellness conditions $p_1^2=m^2$ and $p_2^2=m^2$ making $k=3$. Obviously, this implies that two-loop integrations are not compact \cite{2}, since $\tilde{k}>k$.

The fact that $\tilde{k}<k$ for irreducible three-loop diagrams gives a very strong indication for the (rapid) convergence in the loop expansion. As the numerical results of the next section will show,irreducible three-loop integrations have much more suppressed contributions to the pressure than two-loop integrations for which $\tilde{k} > k$.

\section{Numerics and Results}\label{numres}
\label{chp:numericsandresults}
The following explains the statistical method (Monte-Carlo) in subsection \ref{sec:monte-carlo} which is used then with the constraints emerging from the effective theory (see equations (14), (17) in  \cite{1}) to verify the supports for radial loop integrations in the following subsections \ref{sec:computedratioofpaandpbtop1-loop} and \ref{sec:Theemptyregionofintegrationfordeltapc}. Subsection \ref{sec:twoloopcheck} tests the reliability of the statistical method and compares the shape of the region of radial loop integration for two-loop with the shape of the region of radial loop integration for irreducible three-loop and comes to an important conclusion precisely matching the general discussion in section \ref{chp:constraintsandcompactness}.
\subsection{Monte-Carlo-Integration}
\label{sec:monte-carlo}
The following section describes how the integral of a function $f$  over a region $ G \subset \mathbb{R}^n $ can 
be calculated using statistical methods. The region $G$ is determined by a set of inequalities. Its characteristic 
function will be denoted by $\chi_{G}$. The integral which shall be calculated reads:

\begin{equation}
\label{5_g1}
\int_{G} f(x) dx =\int_{\mathbb{R}^n} \chi_{G}(x)  f(x) dx 
\end {equation}

If $G$ is compact it can be included in a box $B=[x_1,X_1] \times [x_2,X_2] \times ... \times [x_n,X_n]$. 
Therefore the integral can be written:

\begin {equation}
\label{5_g2}
\int_{\mathbb{R}^n} \chi_{G}(x)  f(x) dx =\int_B \chi_{G}(x)  f(x) dx 
\end {equation}

If $B$ has volume $V$, $\frac{1}{V} \chi_B$ can be considered as the probability density function of a 
random variable $X$, which is equally distributed on the box $B$. The integral is exactly the expected value 
$E(V \chi_{G}(X)  f(X) )$:

\begin {equation}
\label{5_g3}
\int_B \chi_{G}(x)  f(x) dx= V \int_B \chi_{G}(x)  f(x) \frac{1}{V} \chi_B dx = V E(\chi_{G}(X) f(X))
\end {equation}

The Monte-Carlo-Method to calculate this integral consists of a statistical estimation of the expected value. It 
is known that the mean value $\bar{X}$  of a sample is an unbiased estimator for the expected value. If one draws 
a random sample $(x_1,x_2...,x_n)$ of points from the box the estimation becomes:

\begin {equation}
\label{5_g4}
\int_{G} f(x) dx \approx V \frac{1}{n}\sum_{i=1}^n \chi_{G}(x_i)  f(x_i)
\end {equation}

The Monte-Carlo-Method is particularly useful to determine integrals over high-dimensional integration regions, 
where deterministic methods would be too time consuming. Unfortunately, the precision of this estimation increases 
only with the root of the sample size. This means that the Monte-Carlo-Method is a fast way to achieve a result with 
a relative precision of about 1 percent, but to achieve one more decimal place in the result the sample size must 
be increased by 100. 

\newpage

\subsection{Computed ratio of $|\Delta P_a|$ \!and $|\Delta P_b|$ to \!$P_{1-loop}$}
\label{sec:computedratioofpaandpbtop1-loop}

The calculation of $\Delta P_a$ and $\Delta P_b$ in \cite{1} includes an integration over a six-dimensional region $G$ with 
variables of integration\footnote{Notice that $x_1$ is integrated analytically through the delta function in  
$\Delta P_a$ and $\Delta P_b$.}  $(x_1,x_2,x_3,z_{12},z_{13},z_{23})$,

\begin {eqnarray}
\label{4_g22}
&& \sum_{l,m,n=1}^2 \!\!\int dx_1 \!\!\int dx_2 \!\!\int dx_3 
\!\!\int dz_{12} \!\!\int dz_{13} \!\!\int_{z_{23,l}}^{z_{23,u}} \!\!\!\!\!\!dz_{23} \times \notag \\
&& f(x_i, z_{ij},\lambda, e) \notag  \\
&& \delta(4e^2+(-1)^{l+m} A(x_1, e)B(x_2,e) -  \\
&& (-1)^{l+n}A(x_1, e) C(x_3, e)- \notag \\
&& (-1)^{m+n}B(x_2, e)C(x_3, e) + ...), \notag
\end {eqnarray}
where A, B and C are square roots $\sqrt{x_i^2 + 4e^2}$ and "..." represents linear terms in $x_i$ and $z_{ij}$ for $i,j = 1, 2, 3$ with $\lambda$ and $e(\lambda)$ as dimensionless temperature and effective gauge coupling, respectively. The whole integrand results from contracting the Feynman rules explained above. The Delta-function is left explicit in order to emphasise its non-trivial and precise integration later.

 As the  $z_{ij}$ stand for cosine values, 
it is possible to restrict their range to the interval $[-1,1]$. The exact shape of $G$ is determined by the (rescaled) compositeness constraints 
\begin{eqnarray}
\label{4_g24}
z_{12} & \leq & \frac{1}{x_1x_2}(4e^2 - A(x_1, e) B(x_2, e) + \frac{1}{2}) \equiv g(x_1, x_2), \notag \\
z_{13} & \geq & \frac{1}{x_1x_3}(-4e^2 + A(x_1, e) C(x_3, e) - \frac{1}{2}) \equiv g(x_1, x_3),  \\
z_{23} & \geq & \frac{1}{x_2x_3}(-4e^2 + B(x_2, e) C(x_3, e) - \frac{1}{2}) \equiv g(x_2, x_3), \notag
\end {eqnarray}

and the additional angular condition on $z_{23, u(l)}$ as defined in \cite{1}.

This angular condition allows a further reduction of the range of $z_{23}$ to the interval $[z_{23,l},z_{23,u}]$. Obviously, this interval depends on the values of $z_{12}$ and $z_{13}$.

The compositeness constraints and the additional angular condition allow a restriction of the $x_i$ to 
the interval $[0,3]$ through considering the support for these radial independent loop integrations which does not 
contradict a nonempty support for the angular integrations \cite{2}.
Therefore a box $B$ which contains $G$ is identified. The compositeness constraints then define the characteristic function  
$\chi_{G}$ whereas the lower and upper limit for $z_{23}$ are used directly as integration limits.

\begin{figure}[!ht]
\begin{center}
\includegraphics[width=10cm, height=5cm]{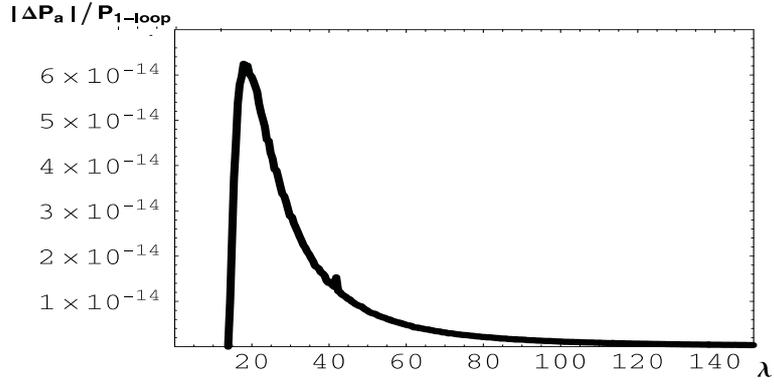}
\caption{Upper estimate for the modulus $|\Delta P_a|/P_{\mbox{1-loop}}$ as a function of $\lambda$ for diagram a).}
\label{5_f1}
\end{center}
\end{figure}

\begin{figure}[!ht]
\begin{center}
\includegraphics[width=10cm, height=5cm]{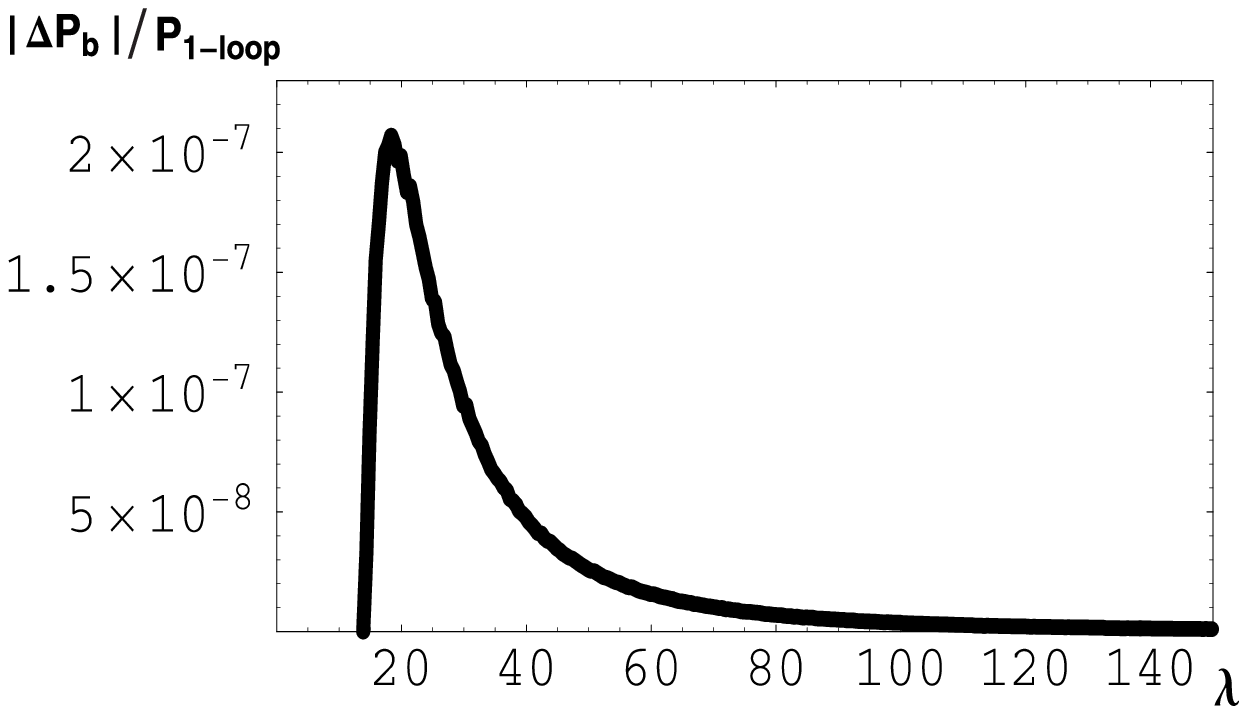}
\caption{Total upper estimate for the modulus $|\Delta P_b|/P_{\mbox{1-loop}}$ as a function of $\lambda$ for diagram b).}
\label{5_f2}
\end{center}
\end{figure}

\begin{figure}[!ht]
\begin{center}
\includegraphics[width=10cm, height=5cm]{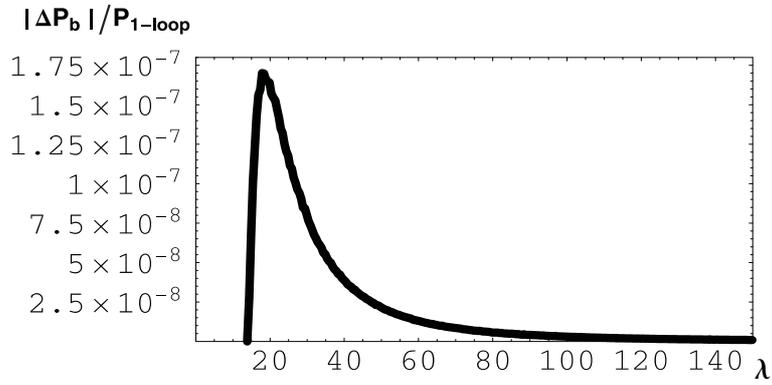}
\caption{Upper estimate for the modulus of the Coulomb part of $|\Delta P_b|/P_{\mbox{1-loop}}$ as a function 
of $\lambda$ for diagram b).}
\label{5_f3}
\end{center}
\end{figure}

\begin{figure}[!ht]
\begin{center}
\includegraphics[width=10cm, height=5cm]{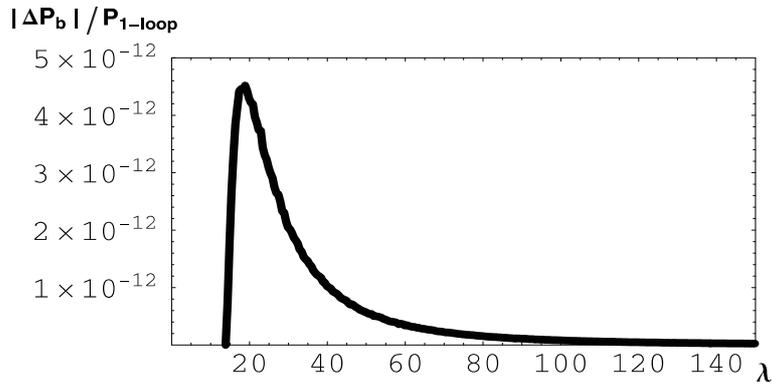}
\caption{Upper estimate for the modulus of the quantum fluctuations of $|\Delta P_b|/P_{\mbox{1-loop}}$ as a 
function of $\lambda$ for diagram b).}
\label{5_f4}
\end{center}
\end{figure}

On the contrary to diagram a) and b) the region of radial loop integration for diagram c) cannot be determined in the 
same way. This has mainly to do with the complexity  and increase in the number of compositeness constraints according to 
additional off-shell variables with arbitrary time components implicated by the four-vertex momentum conservation relation 
of diagram c). Apparently, this makes it impossible to determine the absolute value sign of these compositeness constraints 
from which a small compact region for radial loop integrations could readily follow (respecting  a definite nonempty 
support for angular integrations). The constraints on diagram c) are so restrictive that its region of integration turns 
to be empty (as explained in the next section). 
According to the to the above analysis the ratio of the moduli $\Delta P_a$ and $\Delta P_b$ to one-loop are depicted as a 
function of  the dimensionless temperature $\lambda_c = 13.8 < \lambda < 140$ as the following. The one-loop pressure does not 
include the ground-state contribution. For the effective gauge coupling, the plateau value $e = 8.89$ is used 
for all temperatures $\lambda$ \cite{3}. Throughout most of the deconfining phase, this is admissible, but the logarithmic pole 
of e at the critical temperature is ignored. 
Figure \ref{5_f1} - \ref{5_f4} show that at the critical temperature $\lambda_c = 13.89$ there is no loop contribution 
and $\Delta P_a$ and $\Delta P_b$ approach zero for large temperatures. This is due to the fact that with rising temperature the monopoles become 
massive and dilute, and the scattering processes are suppressed. At asymptotically high temperatures the contributions 
remain finite. The maximum of the ratio of the moduli to one-loop are peaked between $\lambda = 17.5$ and $\lambda = 20$ by $6 \cdot 10^{-14}$
and $2 \cdot 10^{-7}$ for diagram a) and b), respectively. This shows the dominance of diagram b) in the irreducible three-loop 
expansion (the next section shows that diagram c) is vanishing). Comparing figure \ref{5_f2} with \ref{5_f3} and \ref{5_f4}, 
shows the significant dominance of Coulomb fluctuations\footnote{I owe this particular analysis to discussions with Markus Schwarz, see \cite{4,5}}
 $(10^{-7})$ over quantum fluctuation $(10^{-12})$ for the massless propagation in the contributions of diagram b).

\newpage
\vspace{.5cm}
\vspace{.5cm}

\vspace{.5cm}

\vspace{.5cm}
\vspace{.5cm}


\vspace{.5cm}

\clearpage

\subsection{The (empty) region of integration for $|\Delta P_c|$}
\label{sec:Theemptyregionofintegrationfordeltapc}
Consider a box with volume $ 10 \times 10 \times 10 \times 20 \times 2 \times 2 \times 2$ for the loop and angular 
variables $ x_1,x_2,x_3,y_1,z_{12},z_{13},z_{23}$. Through condition $|y_1^2-x_1^2| \leq 1$ in  \cite{1}, and the definition 
of $z_{23,(u,l)}$ a subset of this volume is determined, which roughly represents 2 \% of the volume of the box. In 
this subset $150,\!000,\!000$ points are chosen for $x_1,x_2,x_3,y_1,z_{12},z_{13},z_{23}$ randomly and all four 
possible $\pm$ combinations for $y_2$ and $y_3$ are taken into account. Then it is checked whether the conditions 
(1), (2), (3) and (5) are satisfied which leads to $600,\!000,\!000$ tests. No point satisfies these conditions all 
together which estimates a fraction $1/150,\!000,\!000$ of the subset as the region of integration. This makes a 
fraction by  $1/150,\!000,\!000 \times 0.2$ regarding the box. The box has a volume of $160,\!000$ 
which results a maximal volume of $1/150,\!000,\!000 \times 0.2 \times 160,\!000 \approx 2 \times 10^{-5}$ 
for the region of integration. The typical length distance is 0.2 which means $\sqrt[7]{2 \times 10^{-5}}$. 
It is then highly probable that the region of integration is empty.
A similar analysis is applicable to the vacuum-thermal case showing that its region of integration is also empty. It 
is also possible to argue in terms of compositeness constraints that the region of integration for the vacuum-thermal case 
is automatically empty when the region of integration for vacuum-vacuum is empty \cite{1}.

\subsection{Two-Loop Check}
\label{sec:twoloopcheck}
The Monte-Carlo Method is now used to calculate the results for the local two-loop diagram b) in figure \ref{3_f2}, which 
has already been evaluated analytically \cite{6,7}.
The results are compared in order to check the reliability of the Monte-Carlo Method. The calculation of $\Delta P_b$ of figure 2 in \cite{2,6,7} includes radial loop integration over a 3 dimensional region G with variables of integration $(x,y,z_{xy})$,

\begin{eqnarray}
\label{3_g11}
&& \int dx \int dy \int dz_{xy} f(x, y, z_{xy}, e)\times \\
&& n_B (2\pi\lambda^{-3/2} A(x, e))n_B(2\pi\lambda^{-3/2}B(y, e))\notag 
\end{eqnarray}

with

\begin{equation}
\label{3_g12}
z_{xy} \leq \frac{1}{xy}(4e^2-A(y, e) B(y, e) + \frac{1}{2}) \equiv g(x,y),
\end{equation}

where the Bose factors are left explicit to emphasise their role in radial loop integrations as follows.

In order to carry out this integration with the Monte-Carlo method one needs to bound the region of radial loop integrations by restricting the Bose factors to the following interval :

\vspace{.2cm}
\begin {equation}
\label{5_g5}
0\leq 2\pi \lambda^{-3/2} A(x, e)\leq 10
\end {equation}

and

\begin {equation}
\label{5_g6}
0\leq 2\pi \lambda^{-3/2} B(y, e)\leq 10.
\end {equation}
\vspace{.2cm}

These restrictions ensure that no Boltzmann tails associated with maxi\-mal Boltzmann suppressions are included in 
the region of radial loop integration for diagram b) in figure \ref{3_f2}. The above integrand is plotted in figure \ref{5_f5} below.

\begin{figure}[!ht]
\begin{center}
\includegraphics[scale=0.7]{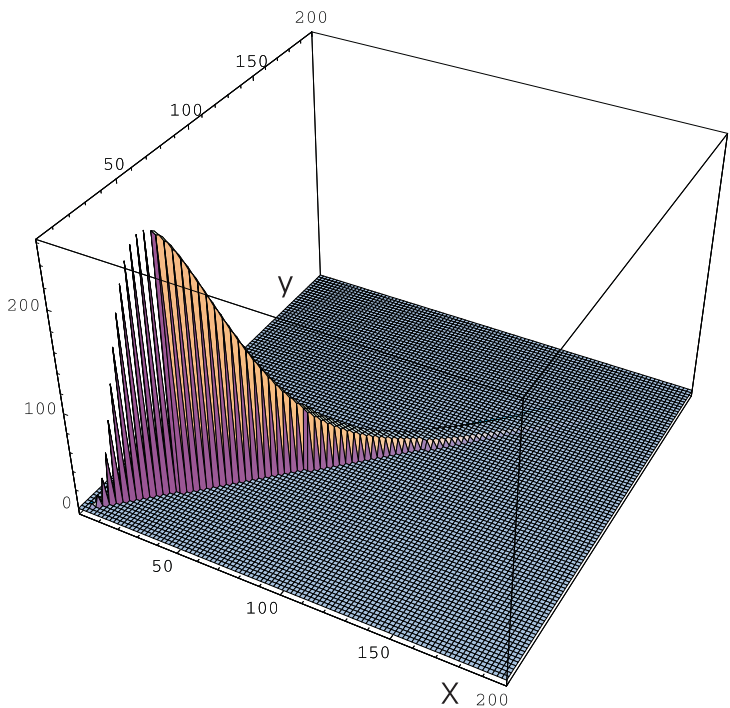}
\caption{The integrand in (\ref{3_g11}) is plotted as a function of $x$ and $y$ for fixed $z_{xy} \equiv \cos\angle(\vec{x},\vec{y})$. The horizontal plane 
represents the x-y-plane, where the integrand (mountain-shaped object) stands on the 'infinite' strip - the domain of definition 
for the contribution in figure \ref{3_f3}\,.}
\label{5_f5}
\end{center}
\end{figure}

In the x-y plane from $x = y = 150$ the maximal suppressions becomes evident which justifies the taken limits (\ref{5_g5}) and 
(\ref{5_g6}). For instance, by taking $x = y = 150, e=8.89$ and $\lambda = 30$, the value of  $2\pi\lambda^{-3/2}A(x, e)$ is $\sim$ 5.77.

This shows that the restrictions in (\ref{5_g5}) and (\ref{5_g6}) are reliable estimations for bounding the region 
of radial loop integration. It should be noticed from figure \ref{5_f1} - \ref{5_f4} and \ref{5_f6} that the upper 
estimate for the contribution, which is of our main interest, takes its values for temperatures $20 \leq \lambda \leq 30$.

\begin{figure}[!ht]
\begin{center}
\includegraphics{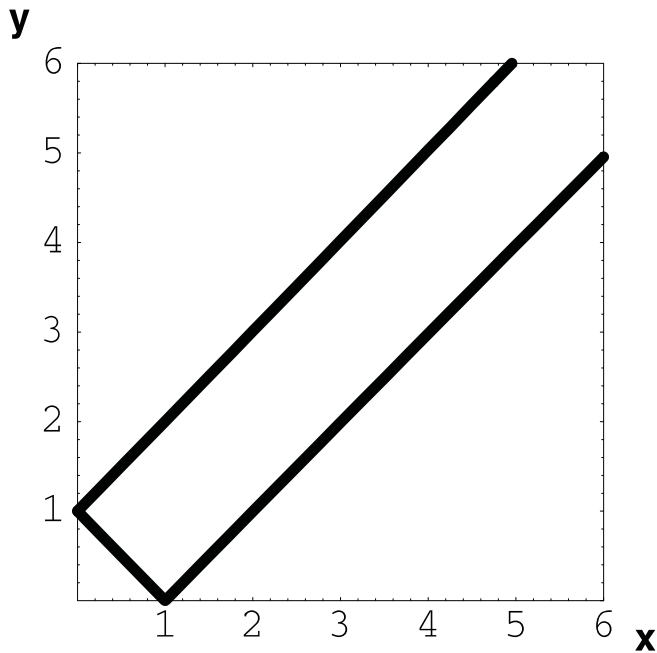}
\caption{Analytical determination of the integration limits for $z_{xy} \equiv \cos\angle(\vec{x}, \vec{y})$ 
depending on $x$ and $y$. In the small triangular region near the origin, the upper limit of integration 
is 1; in the diagonal unbounded strip, the upper limit is $g(x,y)$.  From the two regions adjacent to 
the strip and the axes no contribution arises \cite{7}.}
\label{3_f3}
\end{center}
\end{figure}

\begin{figure}[!ht]
\begin{center}
\includegraphics [width=10cm, height=5cm]{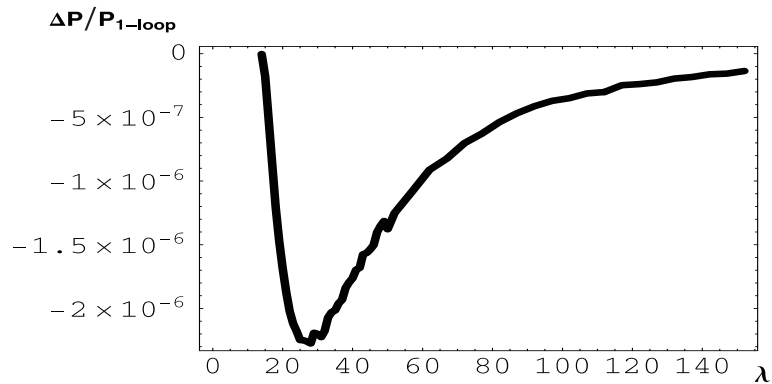}
\caption{The computed ratio of $\Delta P_b$ and $P_{1-loop}$ with the Monte-Carlo method plotted for $13.89 \leq \lambda \leq 140$.}
\label{5_f6}
\end{center}
\end{figure}

There is an analogy between figure \ref{5_f5} and figure \ref{3_f3} in two dimensions. 

The interior region within the infinite strip of figure \ref{5_f5} is exactly the domain of definition of the integrand 
in the x-y plane of figure \ref{3_f3}, where from $x=y=150$ the maximal suppressions become evident.

The ratio of two-loop to one-loop computed by the Monte-Carlo method based on the above restrictions is represented in figure 
\ref{5_f6} below. There are no significant statistical deviations about the analytical results computed in \cite{6,7}. The upper estimate of the analytical 
and statistical results  for $\Delta P/P_{1-loop}$ are both of the order $10^{-6}$ at temperatures\footnote{Notice 
that the plateau value of $e$ in \cite{6,7} is not 8.89 rather than 5.1.} $\lambda = 20 \pm 5$. Obviously this striking compatibility shows the reliability 
and correctness of the Monte-Carlo method for such radial loop integrations. It ensures that the irreducible three-loop results 
represented in this chapter are reliable and correct results. It should be noticed that these irreducible three-loop integrations 
are extremely sensitive and the slightest shift in the calculations can lead to significant changes in the numerical results. For instance, 
one should be aware of the simultaneous $\pm$ sign couplings of the algebraic products within the arguments of the delta 
functions when summing over their zeros in integrations- not all free combinations contribute\footnote{For example, in products containing 
A, B and C terms one should not consider expressions like $\delta (\pm AB \pm AC \pm BC)$ for which there are $2^3 = 8$ free combinations. This is 
because any  choice of sign for A, B and C in 'any one' of the products AB, AC and BC is simultaneously the same choice in the 'other ones' 
(simultaneous sign couplings). To be more specific, if the sign of A in AB is taken negative so it is taken in AC, therefore we should 
consider expressions like $\delta((-1)^{a+b}AB + (-1)^{a+c} AC + (-1)^{b+c} BC)$ with \mbox{$a,b,c = 1,2$} for A, B and C, respectively. The same 
sign coupling condition holds for such products within the unresolved compositeness constraint (17) 
which makes it so restrictive. Taking all combinations free and ignoring the simultaneous sign couplings is including points which 
do not belong to the region of radial loop integration and by that the contributions can be increased significantly. Therefore, the precise Delta Function integration is extra explained in the appendix.}

\vspace{.7cm}

Let us now compare the integrand of (19) in \cite{2} illustrated in figure \ref{5_f5} for the (local) 
two-loop diagram b) in figure \ref{3_f2} with the integrand of (12) in \cite{1} illustrated below in figure \ref{5_f7} 
for the (irreducible) three-loop diagram b) in figure \ref{4_f1}. To get an idea of the 
precise shape of the region of radial loop integration for the three-loop diagram b) in 
figure \ref{4_f1} illustrated in figure \ref{5_f7} the following procedure to reduce the number of dimensions 
has been adopted:

A grid of cosine values for $z_{12}$, $z_{13}$, and $z_{23}$ with width 0.15 is used. This amounts 
to approximatively 2300 combinations of these values. For each of these combinations 
the integrand is calculated as a function of $x_2$ and $x_3$. Then all the diagrams of these 
functions are superposed to get the upper envelope of these diagrams. Figure \ref{5_f7} shows 
this superposition. 
Figure \ref{5_f7} can be interpreted as follows: The area where the superposition (maximum) 
vanishes, does not belong for any combination of the cosine values to the region of 
radial loop integration. So the $x_2-x_3$ extension of this region is at most the area, 
where the plotted upper envelope does not vanish.
It can be seen that figure \ref{5_f7} supports the assertion in \cite{2} that the region of radial loop 
integration is bounded and hence compact. 
The fact that the upper limits of integration 
for $x_2$ and $x_3$ have been set to 3 ensures under all circumstances that the region is 
fully covered\footnote{This is exactly the claim in \cite{2} that the support in $x_2$, $x_3$ 
for the integral in x.yz is contained in the compact set $\{x_2, x_3, < 3\}$ while the support for 
the integration in $z_{12}, z_{13}, z_{23}$ naturally is contained in the set $\{-1 \leq z_{12}, z_{13} \leq +1; z_{23,l} \leq z_{23} \leq z_{23,u}\}$.}.

Recall that Figure \ref{5_f5} shows the integrand for the two-loop diagram for a particular 
choice of the cosine value $z_{xy}$ close to -1. The region of radial loop integration is 
obviously not compact, since the support of the integrand is an 'infinite' strip in the x-y 
plane of figure \ref{3_f3}.
Nonetheless the integrand is suppressed the further $x$ and $y$ are away from the origin.

The remarkable contrast between the two-loop and the (irreducible) three-loop diagram 
is again, as it can be seen from figure \ref{5_f5} and \ref{5_f7}, that the region of 
radial loop integration in the latter case is compact. This was the analytical claim of \cite{2}.

\begin{figure}
\begin{center}
\includegraphics[scale=0.7]{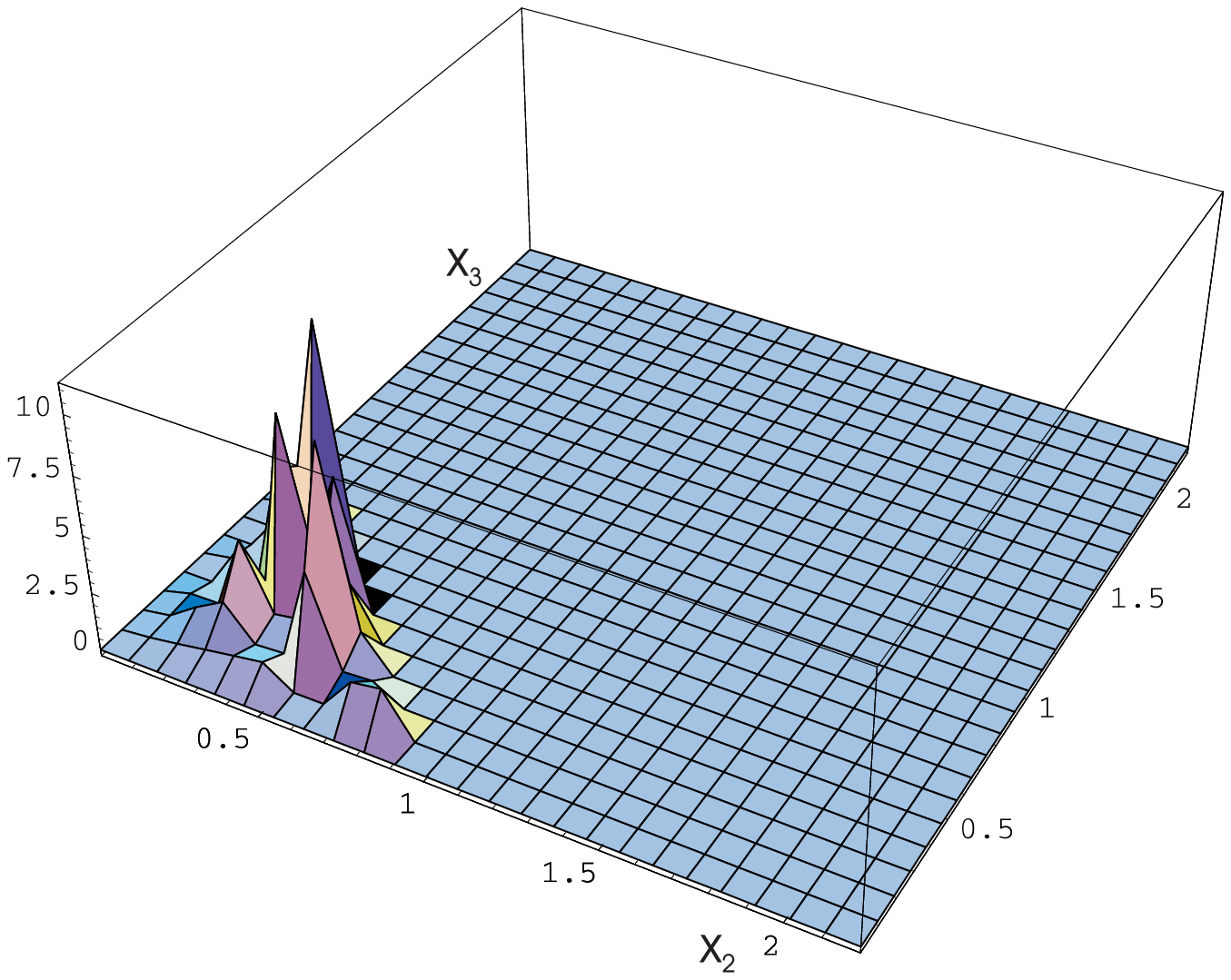}
\caption{The integrand in (\ref{4_g22}) is plotted as a function of $x_2$ and $x_3$ for a grid of 
cosine values \mbox{$z_{12} \equiv \cos \angle (\vec{x_1}, \vec{x_2})$}, \mbox{$z_{13} \equiv \cos \angle 
(\vec{x_1}, \vec{x_3})$} and \mbox{$z_{23} \equiv \cos \angle (\vec{x_2}, \vec{x_3})$}.}
\label{5_f7}
\end{center}
\end{figure}

\newpage
\section{Summary, Conclusions and Outlook}
\label{chp6}

In the present work it was shown that the ramifications due to the increase in the number of 
constraints on the loop momenta of the irreducible three-loop diagrams emerging from 
the effective theory imply either compact or empty supports for integrations and 
accordingly very suppressed contributions to the thermodynamical pressure. This was 
particularly due to the increase in the number of compositeness constraints by the 
s-, t- and u-channels.

The series of extremely suppressed results $10^{-7}$, $10^{-14}$  and $0$ confirm a rapid 
convergence in the loop expansion of SU(2) Yang-Mills thermodynamics. The fact that 
an irreducible three-loop diagram vanishes exactly, where the thermal contribution 
terminates completely, was according to the extremely restrictive compositeness 
constraints that could not be resolved for radial loop integrations implying an 
empty support for integrations.
 
Comparing the modulus of the dominant irreducible three-loop contribution with the 
smallest two-loop contribution reveals that they are nearly compatible, and comparing 
it with the dominant two-loop contribution shows a suppression by a factor $10^{-4}$. 
Apparently, the dominant irreducible three-loop contribution was significantly 
dominated by Coulomb fluctuations over quantum fluctuations for massless propagations. The other nonvanishing 
irreducible three-loop contribution is suppressed by a factor $10^{-11}$ compared to the 
dominant two-loop contribution. The differences between two-loop and irreducible 
three-loop integrations were analysed  according to the relationship between the 
number of independent radial-loop-momentum variables and the number of constraints 
on them. It was shown that the region of radial loop integration for irreducible 
three-loop integrations is, in contrast with the (noncompact) region of radial loop 
integration for two-loop, either compact or empty. This was also shown by the illustration 
of integrands according to the constraints on integrations.

The small numerical results computed for the irreducible three-loop integrations were 
in agreement with the general expectations regarding these diagrams. These numerical 
results were computed by the statistical Monte-Carlo method and this method was explained 
and also tested for two-loop integrations. The compatibility between the statistical 
and the former analytical results for two-loop integrations ensured the reliability 
of the statistical method for the computed irreducible three-loop integrations.

Finally, it could be very interesting to continue with this work by considering higher 
loop diagrams, such as the ones represented below, which can be related to the irreducible 
three-loop diagrams.

The challenge would be to see whether any of them survive the further increase in 
constraints, and hence whether any of them still have infinitesimal contributions at all.
Also, it would important to find a rigorous mathematical proof for the conjecture that the 
dominance of constraints over radial independent loop variables implies compact loop integrations.

\begin{figure}[ht]
\label{6_f1}
\begin{center}
\includegraphics{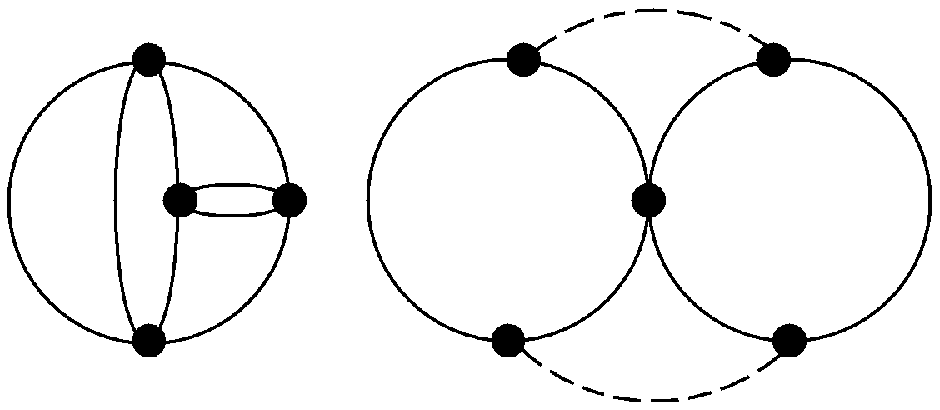}
\end{center}
\end{figure}

\section*{Acknowledgement}

The author thanks Ralf Hofmann for useful discussions and helpful remarks on some points in this paper. The author also thanks Daniel Maitre for help with some technical issues.

\newpage

\begin {appendix}
\section{Delta Function Integration}

The integral over the products of delta functions in giving an 
estimation for the irreducible three-loop contribution of diagram a) and b) reads as follows:

\begin{eqnarray}
\label{a_g1}
\int dx_1&&\!\!\!\!\!\!\!\int dx_2 \int dx_3 \int dy_1 \int dy_2 \int dy_3 \int dz_{12} \int dz_{13}\int dz_{23} \times\notag\\
\notag \\
& &\delta(y_1^2-x_1^2-4e^2) \delta(y_2^2-x_2^2-4e^2)\delta(y_3^2-x_3^2-4e^2) \times\\
& &\delta(4e^2+y_1y_2-x_1x_2z_{12}-(y_1y_3-x_1x_3z_{13})-(y_2y_3-x_2x_3z_{23})) \notag
\end{eqnarray}

The integration will be done numerically after the Dirac delta-functions have been integrated away. The general formula for the 
integration of a delta-function is:

\begin {equation}
\label{a_g2}
\int dx \;\delta(f(x,p))g(x,p) = \sum_{i=1}^{n(p)} {\frac{g(x_i(p), p)}{|\frac{\partial f}{\partial x} (x_i(p),p)|}}
\end {equation}

$n(p)$ denotes the number of zeros of the equation $f(x,p) = 0$. This number depends in general on the parameter $p$. The 
different zeros are then denoted by $x_i(p)$.
To integrate the first delta-function the equation $y_i^2 - x_i^2 - 4e^2 = 0$ has to be solved for $y_1$:

\begin {equation}
\label{a_g3}
y_{1,i} = (-1)^{i} \sqrt{x_1^2+4e^2}
\end {equation}
(\ref{a_g1}) then reads:

\begin {eqnarray}
\label{a_g4}
\int dx_1&&\!\!\!\!\!\!\!\int dx_2 \int dx_3 \int dy_1 \int dy_2 \int dy_3 \int dz_{12} \int dz_{13}\int dz_{23} \notag \\
&&\sum_{i=1}^{2}{\frac{1}{|2(-1)^{i} \sqrt{x_1^2+4e^2}|}}\times\notag \\
&& \delta (y_2^2-x_2^2-4e^2) \delta (y_3^2-x_3^2-4e^2)\times\\
&& \delta (4e^2+(-1)^{i} \sqrt{x_1^2+4e^2} y_2-x_1x_2z_{12}-\notag\\
&& ((-1)^{i} \sqrt{x_1^2+4e^2} y_3-x_1x_3z_{13})-(y_2y_3-x_2x_3z_{23}) \notag
\end {eqnarray}
Proceeding in a similar way with the next two delta-function yields:

\begin {eqnarray}
\label{a_g5}
\int dx_1&&\!\!\!\!\!\!\!\int dx_2 \int dx_3 \int dz_{12}\int dz_{13} \int dz_{23} \notag \\
&&\sum_{i=1}^{2}\sum_{j=1}^{2}\sum_{k=1}^{2} {\frac{1}{8|\sqrt{x_1^2+4e^2}\sqrt{x_2^2+4e^2}\sqrt{x_3^2+4e^2}|}}\times\notag\\
&& \delta (4e^2+(-1)^{i+j} \sqrt{x_1^2+4e^2}\sqrt{x_2^2+4e^2}-x_1x_2z_{12}-\\
&& ((-1)^{i+k} \sqrt{x_1^2+4e^2}\sqrt{x_3^2+4e^2}-x_1x_3z_{13})-\notag \\
&& ((-1)^{j+k} \sqrt{x_2^2+4e^2}\sqrt{x_3^2+4e^2} -x_2x_3z_{23})) \notag
\end {eqnarray}

To remove the remaining delta-function the following equation has to be solved for $x_1$.
One should be aware that squaring an equation is a non-injective operation so that the algebraic solutions need not 
necessarily be solutions to the initial equation.

\begin {eqnarray}
\label{a_g6}
4e^2&+&(-1)^{i+j}\sqrt{x_1^2+4e^2}\sqrt{x_2^2+4e^2}-\notag \\
&&x_1x_2z_{12}-((-1)^{i+k} \sqrt{x_1^2+4e^2}\sqrt{x_3^2+4e^2}-x_1x_3z_{13})-\\
&&((-1)^{j+k} \sqrt{x_2^2+4e^2}\sqrt{x_3^2+4e^2}-x_2x_3z_{23}) = 0 \notag
\end {eqnarray}

\end {appendix}

\end{document}